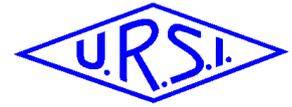

# The technological and scientific development of ASKAP

Bärbel S. Koribalski

Australia Telescope National Facility, CSIRO Astronomy and Space Science, P.O. Box 76, Epping, NSW 1710, Australia

## Abstract

Science results from pilot surveys with the full 36-antenna Australian Square Kilometer Array Pathfinder (ASKAP) have increased strongly over the last few years. This trend is likely to continue with full surveys scheduled to commence later this year. Thanks to novel Phased Array Feeds each ASKAP pointing covers around 30 square degr, making it a fast survey machine delivering high-resolution radio images of the sky. Among recent science highlights are the studies of neutral hydrogen in the Magellanic Clouds as well as nearby galaxy groups and clusters, catalogs of millions of radio continuum sources, the discovery of odd radio circles, and the localization of fast radio bursts, to name just a few. To demonstrate the ASKAP survey speed we also conducted the Rapid ASKAP Continuum Survey (RACS) covering the whole sky south of declination +41 degr at 15 arcsec resolution.

## 1. ASKAP surveys

Pilot Phase II of the large ASKAP science surveys, which were initially approved in 2009, focused on project commensality and is nearly complete. It was preceded by the Early Science Phase (ASKAP-12) and the Pilot Survey Phase I (see Figures 1 & 2). The Pilot Survey Phase II is likely completed by the time this meeting takes place. This will be followed by a consolidation period designed to further improve the telescope systems in advance of full surveys and migration of ASKAP's processing pipeline to the new Pawsey Setonix platform. Once this is done, the full surveys can get under way (pending the outcome of updated proposals submitted in late 2021). For a description of the ASKAP system design and capabilities see Hotan et al. (2021).

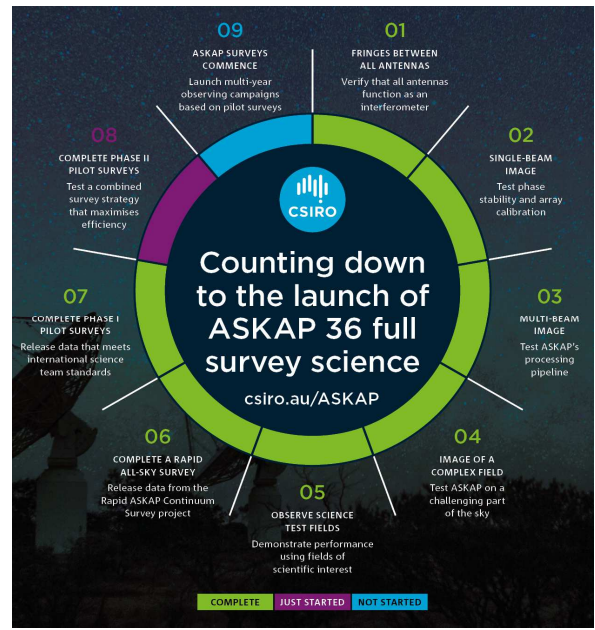

**Figure 2.** The ASKAP count-down clock, created by the ASKAP communications team.

ASKAP data are processed with the ASKAPsoft pipeline (Guzman et al. 2019) at the Pawsey Supercomputing

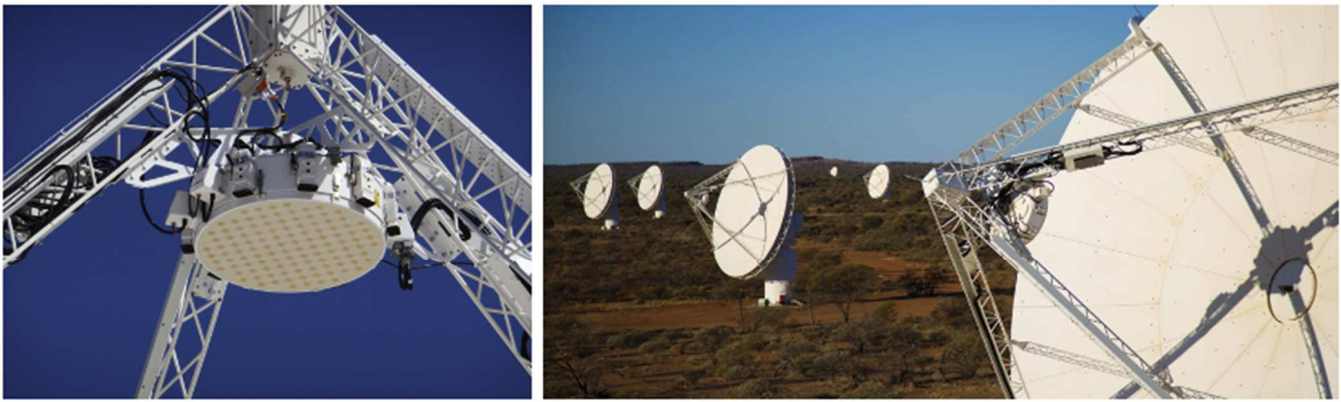

Centre. The pipeline development, which includes the creation of some source catalogues, is carried out in close collaboration with the survey science teams (e.g., Serra et al. 2015a, Heywood et al. 2016, Kleiner et al. 2019). The resulting ASKAP data products are uploaded, validated and made public in the CSIRO ASKAP Science Data Archive (CASDA[1]). An interactive CASDA Skymap[2] allows users to visualize surveys, overlay catalogues and explore the source parameters.

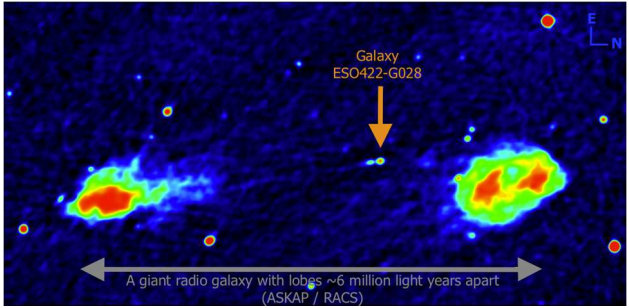

**Figure 3.** ASKAP radio image of the giant radio galaxy ESO 422-G028, obtained from public RACS-low data.

The CASDA Skymap currently highlights the **Rapid ASKAP Continuum Survey (RACS)**, which was conducted as an observatory project to demonstrate the fast survey speed of ASKAP. RACS commenced in April 2019. The RACS-low data (centred at 887.5 MHz) are publicly available, while RACS-mid (1367.5 MHz) and RACS-high (~1700 MHz) observations and processing are under way. For details of the survey design and first results see McConnell et al. (2020). RACS-low (903 fields, 15min integration time per field) has a resolution of ~15 arcsec and an rms median noise of 250 μJy/beam. RACS-mid and RACS-high (1493 fields each, 15 min. integration time per field) are nearing completion.

Another ASKAP observatory project is **SWAG-X**, which covers the GAMA-09 field and the eROSITA Final Equatorial-Depth Survey (eFEDS) at full spectral resolution. It was observed at both 888 and 1296 MHz (together 16h per field) and the first dual-band data (6 of the 12 fields) were released on the 12th of Jan 2022.

There are eight ASKAP survey science projects awaiting time allocations. **WALLABY** (Koribalski et al. 2020) aims to map the 21-cm spectral line of hydrogen (HI) over the southern sky, while also producing deep radio continuum maps as a by-product. It is expected to detect ~0.5 million galaxies, determine their distances, HI and total masses. To search the huge ASKAP spectral data cubes for sources and create reliable source catalogs, the WALLABY team developed a flexible 3D Source Finding Application (SoFiA; Serra et al. 2015b, Westmeier et al. 2021). **EMU** will provide a deep radio continuum survey at slightly lower frequencies (centred on 944 MHz) expecting to detect ~70 million galaxies (Norris et al. 2021a). **FLASH** searches for intervening and associated HI absorption lines against distant bright continuum sources and expects to detect several hundred in each (Allison et al. 2022), while **VAST** aims to detect highly variable and transient radio sources (Murphy et al. 2021). **CRAFT** uses a specially designed mode to detect and localize fast transient radio sources such as FRBs (Bannister et al. 2019). Polarisation studies are conducted by **POSSUM** aiming to produce a catalogue of Faraday rotation measures for around a million extragalactic radio sources (Anderson et al. 2021). **DINGO** is planning to go deep in a small number of pointings, aiming to detect hydrogen in distant galaxies with optical redshifts via HI stacking. **GASKAP** uses high velocity resolution to focus on spectral line studies of the Galaxy Plane and Magellanic Clouds (Pingel et al. 2022).

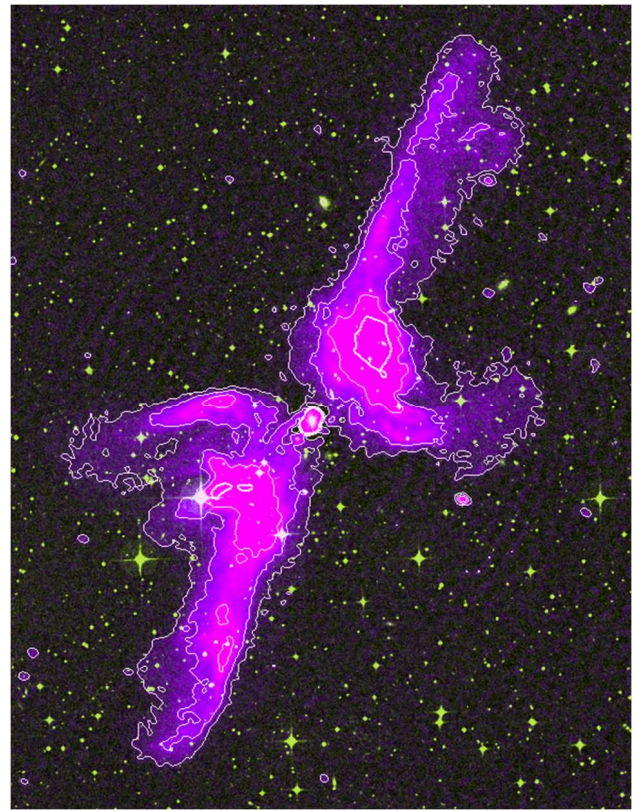

**Figure 4.** ASKAP radio continuum image of the X-shaped radio galaxy PKS 2014-55 from the EMU Pilot Survey (Norris et al. 2021a) overlaid onto an optical image from the Digitized Sky Survey (see also Koribalski 2020).

## 2. Some ASKAP results

Among the multitude of new ASKAP results are some spectacular radio sources. Figure 3 shows an ASKAP continuum image of the giant radio galaxy ESO 422-G028, re-discovered in ASKAP RACS data. The old radio lobes are nearly 2 Mpc apart, and a new inner jet shows its central black hole is active again. Giant radio galaxies stand out because of their large angular sizes and range of morphologies. Figure 4 depicts PKS 2014-55 from the EMU Pilot Survey (Norris et al. 2021a). Its X-shaped radio

(1) https://data.csiro.au/domain/casdaObservation
(2) https://data.csiro.au/domain/casdaSkymap/search

lobes make it appear like a giant butterfly. Much deeper MeerKAT data (Cotton et al. 2020) reveal that backflow is responsible for the radio wings; see also Koribalski (2020).

Figure 5 highlights the nearby galaxy NGC 1371, a member of the gas-rich Eridanus group. WALLABY has been targeting a number of nearby galaxy groups/clusters to investigate the HI content and shape of galaxies in different environments, studying the effects of tidal interactions and ram pressure stripping on their outer disks.

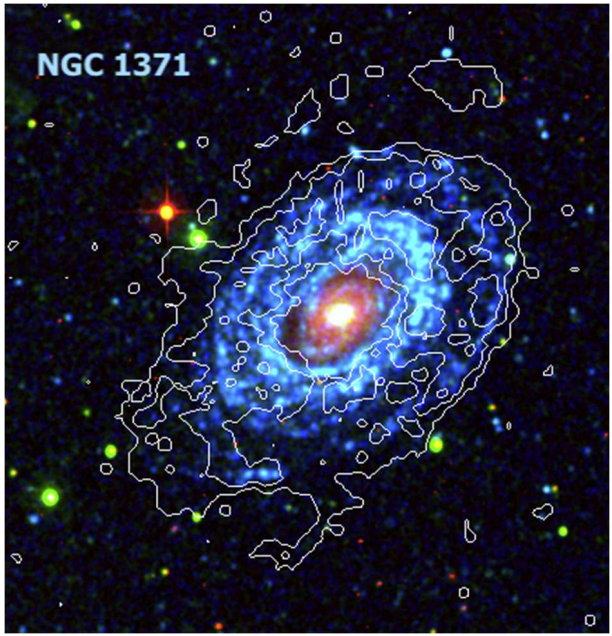

**Figure 5.** ASKAP HI intensity distribution (contours) of the galaxy NGC 1371 from WALLABY, overlaid onto an RGB colour image consisting of GALEX FUV, NUV and DSS B-band images. For details see For et al. (2021).

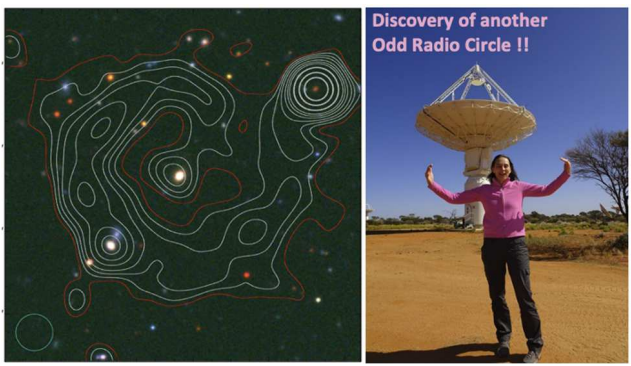

**Figure 6.** *Left:* ASKAP radio continuum image (contours) of the recently discovered odd radio circle ORC J0102-2450 overlaid onto a 3-color optical image from the Dark Energy Survey (DES). The 13 arcsec ASKAP beam is shown in the bottom left. For details see Koribalski et al. (2021). *Right:* Visiting ASKAP and the Murchison Radio-Astronomy Observatory with one of the 36 ASKAP antennas in the background.

(1) https://data.csiro.au/domain/casdaObservation
(2) https://data.csiro.au/domain/casdaSkymap/search

Odd radio circles (ORCs) were first discovered by Norris et al. (2021b) in the EMU Pilot Survey. A recent ASKAP discovery, ORC J0102-2450 (see Figure 6), was the third such radio circle with a prominent central galaxy (Koribalski et al. 2021), suggesting typical diameters of 300 – 500 kpc. None of the radio rings have – as yet – counterparts at any other wavelength. While some ideas have been put forward, likely related to an energetic event in the associated central galaxy, the ORC formation mechanism currently remains unknown. Deep MeerKAT follow-up observations are in hand and analysis is on-going. As numerical simulations are under way to explore possible ORC origins, the search for more single and double ORCs continues.

Double radio relics, like the pair shown in Figure 7, highlight shock fronts in galaxy clusters, which tend to be filled with hot gas detectable in X-ray. This makes for a very strong synergy between radio, X-ray and optical data to study galaxy cluster formation, dynamics and evolution (e.g., Brüggen et al. 2021).

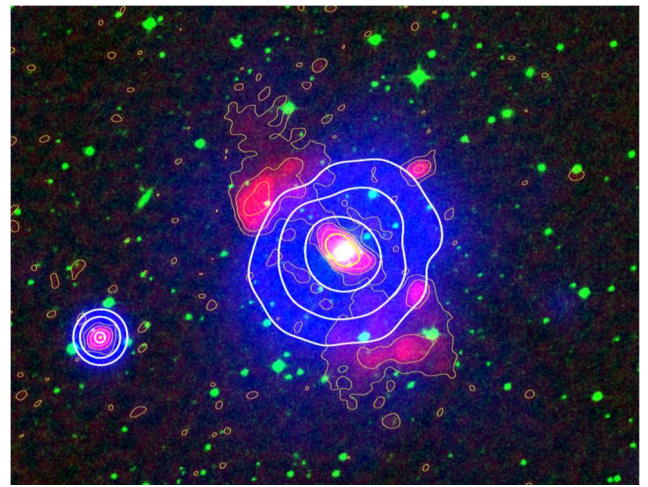

**Figure 7.** RGB colour image of the RXC J2143.9-5637 galaxy cluster (redshift z = 0.0824), consisting of ASKAP radio continuum emission (red colours and yellow contours) from the EMU Pilot Survey (Norris et al. 2021a), XMM-Newton X-ray (blue colours and white contours), and DSS optical (green colours) data.

The upcoming full ASKAP surveys will provide a flood of data (spectral line, radio continuum, transients, …) publicly available through CASDA. As a Pathfinder to the Square Kilometer Array, ASKAP is providing us with many valuable technology, computing and science lessons. ASKAP has huge discovery potential, often realized by comparison of the source radio emission with a wide range of multi-wavelength surveys and close collaborations within the international community.

## 6. Acknowledgements

ASKAP is part of the Australia Telescope National Facility (ATNF) which is managed by CSIRO. Operation of ASKAP is funded by the Australian Government with support from the National Collaborative Research Infrastructure Strategy. ASKAP uses the resources of the Pawsey Supercomputing Centre. Establishment of ASKAP, the Murchison Radio-astronomy Observatory and the Pawsey Supercomputing Centre are initiatives of the Australian Government, with support from the Government of Western Australia and the Science and Industry Endowment Fund.

We acknowledge the Wajarri Yamatji people as the traditional owners of the Observatory site.

---

(1) https://data.csiro.au/domain/casdaObservation
(2) https://data.csiro.au/domain/casdaSkymap/search